\documentstyle [12pt,epsf] {article}

\begin{document}
\begin{titlepage}

\vspace{2cm}
\Large
{\bf Stellar masses, star formation rates, metallicities and
AGN properties  for $2 \times 10^5$ galaxies  in the SDSS 
Data Release Two (DR2)}
\\                                                     
\begin{center}
\large
Jarle Brinchmann$^{1,2}$, St\'ephane Charlot$^{1,3}$, Timothy M. Heckman$^4$,\\
Guinevere Kauffmann $^{1}$, Christy Tremonti$^{5,4}$,  Simon D.M. White$^1$  
\\
\end{center}
\vspace{0.8cm}
\begin {abstract}
By providing homogeneous photometric and spectroscopic data of high quality
for very large and objectively selected samples of galaxies, the Sloan Digital
Sky Survey allows statistical studies of the physical properties of galaxies
and AGN to be carried out at an unprecedented level of precision and
detail. Here we publicly release catalogues of derived physical properties for
211,894 galaxies, including 33,589 narrow-line AGN. These are complete samples with
well understood selection criteria drawn from the normal galaxy spectroscopic
sample in the second SDSS public data release (DR2). We list properties
obtained from the SDSS spectroscopy and photometry using modelling techniques
presented in papers already published by our group.  These properties include:
stellar masses; stellar mass-to-light ratios; attenuation of the starlight by
dust; indicators of recent major starbursts; current total
and specific star-formation rates, both for the regions with spectroscopy and
for the galaxies as a whole; gas-phase metallicities; AGN classifications
based on the standard emission line ratio diagnostic diagrams and  AGN 
[OIII] emission line luminosities. We also list our            
own measurements of absorption line indices and emission line fluxes from
which these quantities were derived, together with a few quantities from the
standard SDSS pipelines which play an important role in our work.  Many other
observed properties of these galaxies can be obtained from the SDSS DR2
catalogues themselves. We will add further physical properties to this release
site as the relevant papers are accepted for publication.
Catalogues containing these parameters may be accessed at
http://www.mpa-garching.mpg.de/SDSS/.
\normalsize
\end {abstract}
\vspace {0.8 cm}
\normalsize
\vspace{0.7cm} 
\small
{\em $^1$Max-Planck Institut f\"{u}r Astrophysik, D-85748 Garching, Germany} \\
{\em $^2$ Centro de Astrof{\'\i}sica da Universidade do Porto, 
Rua das Estrelas - 4150-762 Porto, Portugal}\\
{\em $^3$ Institut d'Astrophysique du CNRS, 98 bis Boulevard Arago, F-75014 Paris, France} \\
{\em $^4$Department of Physics and Astronomy, Johns Hopkins University, Baltimore, MD 21218}\\
{\em $^5$ Steward Observatory, 933 N.\ Cherry Ave., Tucson, AZ
  85721, USA}\\
\end {titlepage}
\normalsize

\section{The SDSS Data}

The galaxy spectra we have analyzed are drawn from the Sloan Digital Sky
Survey (SDSS).  The survey goals are to obtain photometry of a quarter
of the sky and spectra of nearly one million objects.  Imaging is
obtained in the \emph{u, g, r, i, z} bands (Fukugita et al 1996;
Smith et al 2002) with a special purpose drift scan camera
(Gunn et al 1998) mounted on the SDSS 2.5~meter telescope at
Apache Point Observatory.  The imaging data are photometrically
(Hogg et al 2001) and astrometrically (Pier et al 2003)
calibrated, and used to select stars, galaxies, and quasars for
follow-up fiber spectroscopy.  Spectroscopic fibers are assigned to
objects on the sky using an efficient tiling algorithm designed to
optimize completeness (Blanton et al 2003).  The details of the
survey strategy can be found in (York et al 2000) and an
overview of the data pipelines and products is provided in the Early
Data Release paper (Stoughton et al 2002).

Our sample for this study is composed of 211,894 objects which
have been spectroscopically confirmed as galaxies and have data
publicly available in the SDSS Data Release~2
(Abazajian et al 2004).  These galaxies are part of the
SDSS `main' galaxy sample used for large scale structure studies
(Strauss et al 2002) and have Petrosian $r$ magnitudes in the
range $14.5 < r < 17.77$ after correction for foreground galactic
extinction using the reddening maps of
Schlegel,Finkbeiner \& Davis (1998).  Their redshift
distribution extends from $\sim0.005$ to 0.30, with a median $z$ of 0.10.

The spectra are
obtained with two 320-fiber spectrographs mounted on the SDSS 
2.5-meter telescope.  Fibers 3 arcseconds in diameter are manually plugged
into custom-drilled aluminum plates mounted at the focal plane of the
telescope. The spectra are exposed for 45 minutes or until a fiducial
signal-to-noise (S/N) is reached.  The median S/N per pixel for
galaxies in the main sample is $\sim14$.  The spectra are processed by
an automated pipeline (Schlegel et al., in prep.)  which flux and
wavelength calibrates the data from 3800 to 9200~\AA.  The
instrumental resolution is R~$\equiv \lambda/\delta\lambda$ = 1850 --
2200 (FWHM$\sim2.4$~\AA\ at 5000~\AA).

The Survey has been able to obtain a remarkable level of
spectrophotometric precision by the simple practice of observing
multiple standard stars simultaneously with the science targets.  (The
artifice in this case is that the `standards' are not classical
spectrophotometric standards, but are halo F-subdwarfs that are
calibrated to stellar models -- see Abazajian et al (2004) for
details.)  To quantify the quality of the spectrophotometry we have
compared magnitudes synthesized from the spectra with SDSS photometry
obtained with an aperture matched to the fiber size. The 1$\sigma$
error in the synthetic colors is 5\% in $g-r$ and 3\% in $r-i$
($\lambda_{g}\sim4700$~\AA; $\lambda_{r}\sim6200$~\AA;
$\lambda_{i}\sim7500~$\AA).  At the bluest wavelengths
($\sim3800$~\AA) we estimate the error to be $\sim12$\% based on
repeat observations.  

\section{Emission Line Measurements}

The optical spectra of galaxies are very rich in stellar \emph{absorption}
features, which can complicate the measurement of nebular emission
lines. In order to maintain speed and flexibility, the SDSS
spectroscopic pipeline performs a very simple estimate of the stellar
continuum using a sliding median.  While this is generally adequate
for strong emission lines, a more sophisticated treatment of the
continuum is required to recover weak features and to properly account
for the stellar Balmer absorption which can reach equivalent widths of
5~\AA\ in some galaxies.  To address this need, we have designed a
special-purpose code optimized for use with SDSS galaxy spectra which
fits a stellar population model to the continuum.  We adopt the basic
assumption that any galaxy star formation history can be approximated
as a sum of discrete bursts.  Our library of template spectra is
composed of single stellar population models generated using the new
population synthesis code of Bruzual \& Charlot (2003; BC03).
The BC03 models incorporate an
empirical spectral library (Le Borgne et al 2003; in preparation) with a
wavelength coverage (3200 - 9300 \AA) and spectral resolution
($\sim3$~\AA) which is well matched to that of the SDSS data.  Our
templates include models of ten different ages (0.005, 0.025, 0.1,
0.2, 0.6, 0.9, 1.4, 2.5, 5, 10 Gyr) and three metallicities (1/5
$Z_{\odot}$, $Z_{\odot}$, and 2.5 $Z_{\odot}$).  For each galaxy we
transform the templates to the appropriate redshift and velocity
dispersion and resample them to match the data.  To construct the best
fitting model we perform a non-negative least squares fit with dust
attenuation modeled as an additional free parameter.  In practice, our
ability to simultaneously recover ages and metallicities is strongly
limited by the signal-to-noise of the data.  Hence we model galaxies as
single metallicity populations and select the metallicity which
yields the minimum $\chi^2$.  
The details of the template fitting code
will be presented in Tremonti et al.~ (in prep.).  

After subtracting the best-fitting stellar population model of the
continuum, we remove any remaining residuals (usually of order a few 
percent) with a sliding 200 pixel median, and fit the nebular emission lines.
Since we are interested in recovering very weak nebular features, we adopt a
special strategy: we fit all the emission lines with Gaussians
simultaneously, requiring that all of the Balmer lines
(H$\delta$, H$\gamma$, H$\beta$, and H$\alpha$) have the same line
width and velocity offset, and likewise for the forbidden lines
([OII]~$\lambda\lambda 3726, 3729$, 
[OIII]~$\lambda\lambda4959, 5007$,
[NII]~ $\lambda\lambda6548, 6584$, 
[SII]~$\lambda\lambda6717, 6731$).  
We are careful to take into account the
wavelength-dependent instrumental resolution of each fiber, which is
measured for each set of observations by the SDSS spectroscopic
pipeline from the arc lamp images.  The virtue of constraining the
line widths and velocity offsets is that it minimizes the number of
free parameters and effectively allows the stronger lines to be used
to help constrain the weaker ones.  Extensive by-eye inspection
suggests that our continuum and line fitting methods work well.

\section {Parameters derived from Spectra}

\subsection {Stellar masses}

{\bf References:\\ 
Kauffmann, G. et al., 2003, MNRAS, 341,33;\\
Kauffmann, G. et al., 2003, MNRAS, 341, 54}\\ 
We developed a method to constrain the star formation histories, dust 
attenuation and stellar masses of galaxies.
It is based on two stellar absorption line indices, the 4000 \AA \hspace{0.1cm}
break strength and the Balmer absorption line index H$\delta_A$.
Together, these indices allow us to constrain the mean stellar ages  
of galaxies and the fractional stellar mass formed in bursts over the past few Gyr.
A comparison with broad band photometry then yields estimates of dust attenuation
and of stellar mass.
The stellar mass catalogues include the 4000 \AA\ break strength and H$\delta_A$
measurements, stellar mass estimates, mass-to-light ratios, estimates of
dust attenuation, and estimates of the fraction of the stellar
mass formed in bursts in the past 2 Gyr.

\subsection {Star formation rates} 

{\bf Reference:\\ 
Brinchmann, J. et al., 2004, MNRAS, in press (astro-ph/0311060)}\\
Our methods for deriving star formation rates  inside the
fibre aperture  make use of the 
methodology described in Charlot et al (2002) and emission line models
described in Charlot \& Longhetti (2001; CL01).
The CL01 model is based on a combination of
the Bruzual \& Charlot (1993)  and Ferland (1996, version
C90.04) population synthesis and photoionization codes. In the model,
the contributions to the nebular emission by HII  regions and
diffuse ionized gas are combined and described in terms of an
effective (i.e.\ galaxy-averaged) metallicity, ionization parameter,
dust attenuation at 5500~\AA, and dust-to-metal ratio.  The depletion
of heavy elements onto dust grains and the absorption of ionizing
photons by dust are included in a self-consistent way.
We have developed a method to aperture correct our star formation rates  
using resolved imaging and we have shown that our method takes out
essentially all aperture bias in the star formation rate (SFR)
estimates, allowing an accurate estimate of the total SFRs in
galaxies. Our catalogues also include estimates of the specific  
star formation rates of the DR2 galaxies (the star formation rate per
unit stellar mass) both inside the fiber and for the galaxy as a whole.

\subsection {Gas-phase metallicities}

{\bf Reference:\\ 
Tremonti, C. et al., 2004, ApJ, in press (astro-ph/0405537)}\\
Gas-phase metallicities have been measured using the approach outlined in 
Charlot et al (in prep.). 
The metallicity estimates are based on simultaneous fits of
all the most prominent emission lines ([OII], H$\beta$,
[OIII], [HeI], [OI], H$\alpha$, [NII],
[SII]) using the CL01 models.                              
In the paper referenced above, we compare our                                 
derived metallicities with estimates based on the
ratio R$_{23}$ = ([OII] + [OIII])/H$\beta$,
and show that our estimates are similar 
to  previous strong-line calibrations. 

\subsection {AGN catalogue}

{\bf References:\\
Kauffmann, G. et al , 2003, MNRAS, 346, 1055;\\ 
Heckman, T.M. 
et al, 2004, ApJ, in press }\\
AGN have been selected on the basis of their position on the BPT
(Baldwin, Phillips \& Terlevich 1981) diagram. We 
select AGN using  the ratio
[OIII]$\lambda$5007/H$\beta$ versus the ratio [NII]/H$\alpha$ for all galaxies
where all
four lines were detected with $S/N>3$. 
In the studies referenced above, the luminosity of the [OIII] emission line
was used as a tracer
of the strength of activity in the nucleus and these measurements are included
in the catalogue along with the stellar velocity dispersions, which were used
as an estimate of the black hole mass.

\subsection {Emission line fluxes}

In addition to our physical parameter estimates, we include a table of the
measured emission line fluxes used in our work. These include
[OII]$\lambda$3726,$\lambda$3729, H$\beta$, [OIII]$\lambda$5007, 
H$\alpha$, [NII]$\lambda$6584 and [SII]$\lambda$6717.

\vspace{1.5cm}

\large
{\bf Acknowledgements}\\

\normalsize
The data processing
software developed for this project benefited from IDL routines
written by Craig Markwardt and from the IDL Astronomy User's Library
maintained by Wayne Landsman at Goddard Space Flight Center.
C.~A.~T. thanks the Max Planck Institute for Astrophysics and the
Johns Hopkins Center for Astrophysics for their generous financial
support.  She also acknowledges support from NASA grant NAG~58426 and
NSF grant AST-0307386.  J.~B. and S.~C. thank the Alexander von
Humbolt Foundation, the Federal Ministry of Education and Research,
and the Programme for Investment in the Future (ZIP) of the German
Government for their support. J.~B. would also like to acknowledge the
receipt of an ESA post-doctoral fellowship.

Funding for the creation and distribution of the SDSS Archive has been
provided by the Alfred P. Sloan Foundation, the Participating
Institutions, the National Aeronautics and Space Administration, the
National Science Foundation, the U.S. Department of Energy, the
Japanese Monbukagakusho, and the Max Planck Society. The SDSS Web site
is http://www.sdss.org/.
The SDSS is managed by the Astrophysical Research Consortium (ARC) for
the Participating Institutions. The Participating Institutions are The
University of Chicago, Fermilab, the Institute for Advanced Study, the
Japan Participation Group, The Johns Hopkins University, Los Alamos
National Laboratory, the Max-Planck-Institute for Astronomy (MPIA),
the Max-Planck-Institute for Astrophysics (MPA), New Mexico State
University, University of Pittsburgh, Princeton University, the United
States Naval Observatory, and the University of Washington.

\pagebreak
\Large
\begin {center} {\bf References} \\
\end {center}
\normalsize
\parindent -7mm
\parskip 3mm

Abazaijian, K. et al, 2004, AJ, in press

Baldwin, J., Phillips, M., \& Terlevich, R. 1981, PASP, 93, 5

Blanton, M.R., Lin, H., Lupton, R.H., Maley, F.M., Young, N., Zehavi, I.,
Loveday, J., 2003, AJ, 125, 2276

Brinchmann, J. et al., 2004, MNRAS, in press (astro-ph/0311060)

Bruzual, G., \& Charlot, S. 2003, MNRAS, 344, 1000

Charlot, S., Longhetti, M., 2001, MNRAS, 323, 887

Charlot, S., Kauffmann, G., Longhetti, M., Tresse, L., White, S.D.M.,
Maddox, S.J., Fall, S.M., 2002, MNRAS, 330, 876

Ferland, G., 1996, Hazy, A Brief Introduction to Cloudy. Internal Report,
University of Kentucky, USA 

Fukugita, M., Ichikawa, T., Gunn, J.E., Doi, M., Shimasaku, K., 
Schneider, D.P. 1996, AJ, 111, 1748

Gunn, J. et al, 1998, ApJ, 116, 3040                                   

Heckman, T.M. et al, 2004, ApJ, in press

Hogg, D., Finkbeiner, D., Schlegel, D., \& Gunn, J. 2001, AJ, 122, 2129

Kauffmann, G. et al, 2003, MNRAS, 341, 33                                
                           
Kauffmann, G. et al, 2003, MNRAS, 341, 54                                 

Kauffmann, G. et al , 2003, MNRAS, 346, 1055 

Pier, J., et al 2002 AJ, in press

Schlegel, D.J, Finkbeiner, D.P., Davis, M., 1998, ApJ, 500, 525

Smith, J.A. et al, 2002, AJ, 123, 2121

Stoughton, C. et al, 2002, AJ, 123, 485                                    

Strauss, M. et al, 2002, AJ, 124, 1810                                      

Tremonti, C. et al., 2004, ApJ, in press (astro-ph/0405537)

York, D.G. et al. 2000, AJ, 120, 1579

\end{document}